\documentclass[11pt]{article}

\usepackage[final]{acl}

\usepackage{times}
\usepackage{latexsym}

\usepackage[T1]{fontenc}

\usepackage[utf8]{inputenc}

\usepackage{microtype}

\usepackage{inconsolata}

\usepackage{graphicx}

\usepackage{amsmath}
\usepackage{amssymb}
\usepackage{subcaption}
\usepackage{booktabs}
\usepackage{multirow}
\usepackage{array}
\usepackage{tcolorbox}

\definecolor{tmp}{RGB}{0, 102, 204}

\title{\includegraphics[height=0.8em]{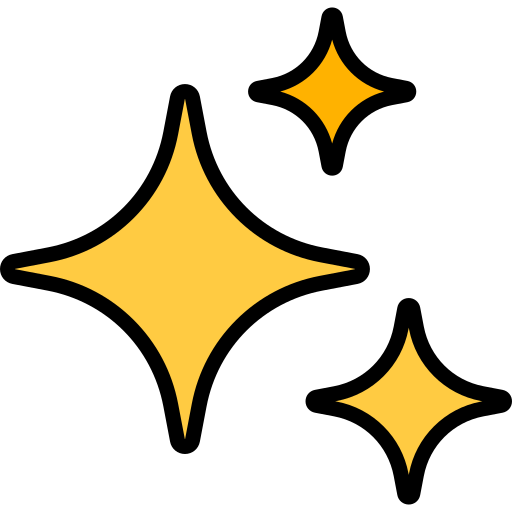}\textsc{ArcLight}: A \textsc{Light}weight LLM Inference \textsc{Arc}hitecture\\for Many-Core CPUs}

\author{Yuzhuang Xu$^{1}$ \quad Xu Han$^{2,*}$ \quad Yuxuan Li$^{2}$ \quad Wanxiang Che$^{1,}$\thanks{Corresponding authors} \\
\textsuperscript{\rm 1}Harbin Institute of Technology, Harbin, China \\
\textsuperscript{\rm 2}Tsinghua University, Beijing, China \\
\texttt{\{xyz, car\}@ir.hit.edu.cn, han-xu@mail.tsinghua.edu.cn} }

\begin{document}
\maketitle
\begin{abstract}
Although existing frameworks for large language model (LLM) inference on CPUs are mature, they fail to fully exploit the computation potential of many-core CPU platforms. Many-core CPUs are widely deployed in web servers and high-end networking devices, and are typically organized into multiple NUMA nodes that group cores and memory. Current frameworks largely overlook the substantial overhead of cross-NUMA memory access, limiting inference scalability and intelligence enabling on such platforms. 
To address this limitation, we build \textsc{ArcLight}, a lightweight LLM inference architecture designed from the ground up for many-core CPUs. \textsc{ArcLight} integrates efficient memory management and thread scheduling, and introduces finely controlled tensor parallelism to mitigate the cross-node memory access wall. Experimental results show that \textsc{ArcLight} significantly surpasses the performance ceiling of mainstream frameworks, achieving up to 46\% higher inference throughput. Moreover, \textsc{ArcLight} maintains compatibility with arbitrary CPU devices. \textsc{ArcLight} is publicly available at \url{https://github.com/OpenBMB/ArcLight}.
\end{abstract}

\section{Introduction}

The rapid proliferation of large language models (LLMs) sparks a significant demand for efficient inference frameworks. 
Currently, this area is dominated by two paradigms, high-performance GPU-based systems~\citep{efficient2023,sglang2024} and accessible CPU-based systems~\citep{llamacpp2023}. 
Among the latter, \textit{llama.cpp}~\citep{llamacpp2023} stands out as a pioneering framework specifically designed for CPU inference, enabling LLM inference through an efficient C++ implementation.

\begin{figure}[t]
  \centering
  \includegraphics[width=0.92\columnwidth]{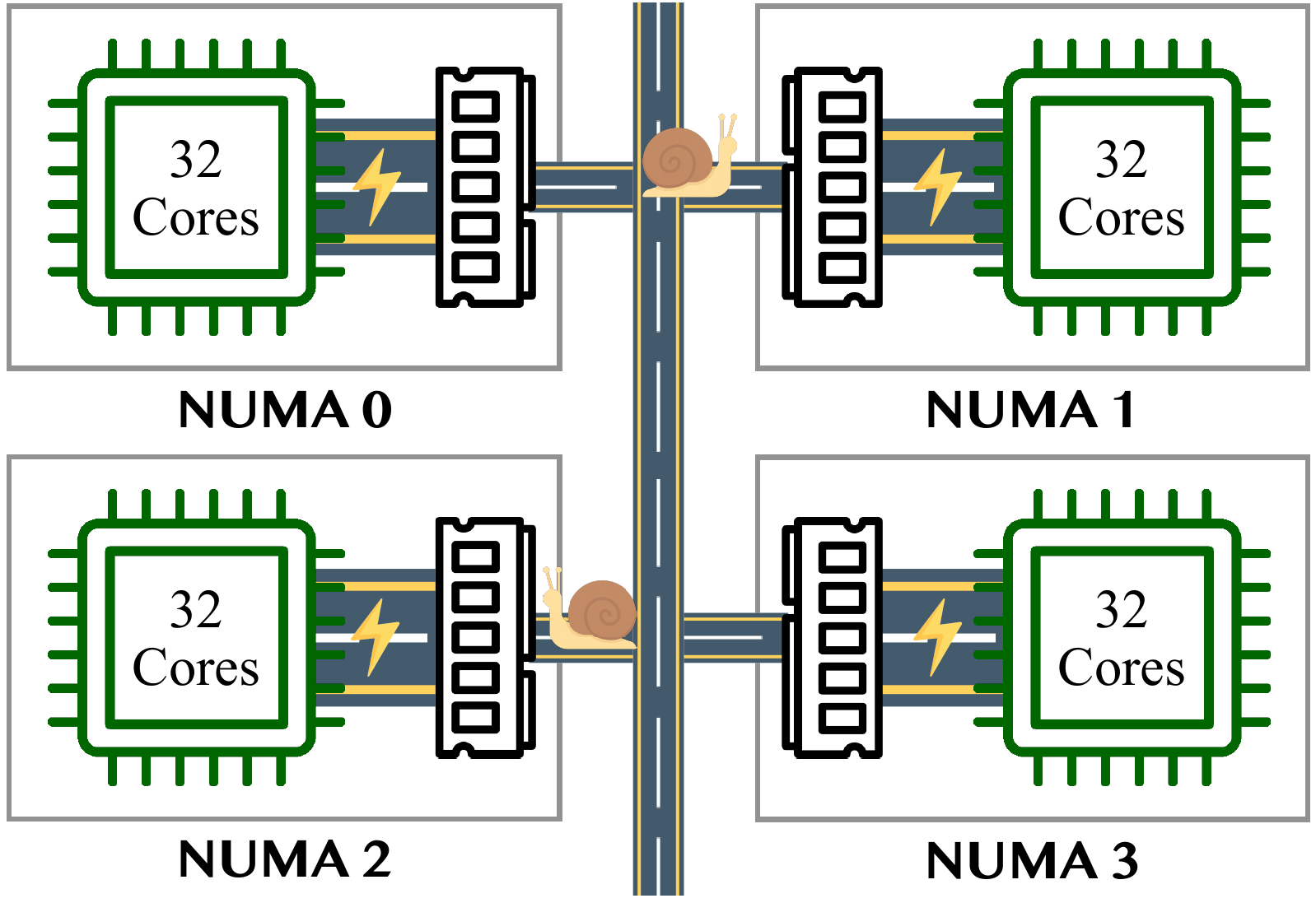}
  \caption{For many-core CPU platforms, CPU cores and memory are organized using NUMA nodes. Here shows a platform with 128 cores and 4 NUMA nodes.}
  \label{fig:intro}
  \vspace{-5mm}
\end{figure}

However, mainstream CPU frameworks struggle to adapt to many-core CPU platforms, which are common in web servers and high-end networking devices. These many-core CPU platforms typically adopt a Non-Uniform Memory Access (NUMA) architecture, in which CPU cores and memory are partitioned across multiple nodes. For the NUMA architecture, the cost of accessing remote memory (across nodes) is significantly higher than local access, as shown in Figure~\ref{fig:intro}. Existing frameworks often overlook this ``cross-NUMA memory access wall''. When scaling LLM inference to dozens of cores, the overhead of data synchronization and non-local memory access becomes a bottleneck, preventing the system from reaching its theoretical computational ceiling.

Unfortunately, retrofitting existing frameworks to handle many-core challenges is a daunting task. Adapting a mature system like \textit{llama.cpp} for NUMA-aware efficiency requires ``surgical'' refactoring of the entire stack, from low-level memory allocation and thread management to high-level model definition. Furthermore, as mainstream frameworks evolve, they tend to become increasingly bloated. The accumulation of legacy code and complex abstractions makes the internal logic opaque, hindering researchers from implementing fine-grained optimizations or quickly supporting emerging model architectures.

To bridge this gap, we present \textsc{ArcLight}, a lightweight LLM inference architecture built from the ground up for many-core CPUs. \textsc{ArcLight} is guided by a philosophy of minimalism and modularity, containing only the essential components required for high-performance inference. Its clear structural boundaries allow developers to easily extend functionality or integrate new models without navigating a monolithic codebase. More importantly, \textsc{ArcLight} introduces targeted optimizations for many-core environments, including NUMA-aware memory and thread management and finely controlled tensor parallelism to mitigate cross-node memory access overhead.

Our experimental results demonstrate that \textsc{ArcLight} achieves up to 46\% higher inference throughput compared to the currently popular \textit{llama.cpp} on many-core CPU platforms. The primary contributions of this work are two-fold:
\begin{itemize}
    \item \textbf{A Lightweight Inference Architecture:} We introduce and open-source a modular framework that distills LLM inference to its core essentials. This provides a transparent, ``hackable'' foundation for researchers to experiment with CPU-based LLM deployment without the overhead of traditional frameworks.

    \item \textbf{Optimization for Many-Core CPUs:} We provide a blueprint for multi-dimensional optimization on many-core CPUs. By addressing thread scheduling and the memory access wall, we significantly push the performance boundaries of CPU-based LLM inference.
\end{itemize}

\section{System Design}

\textsc{ArcLight} follows a clean and well-defined modular design, implementing the core features of CPU-based inference in a minimalist manner to facilitate future extensions and iterations. In this section, we present the overall architecture of \textsc{ArcLight} and describe its key designs.

\begin{figure}[t]
  \centering
  \includegraphics[width=0.95\columnwidth]{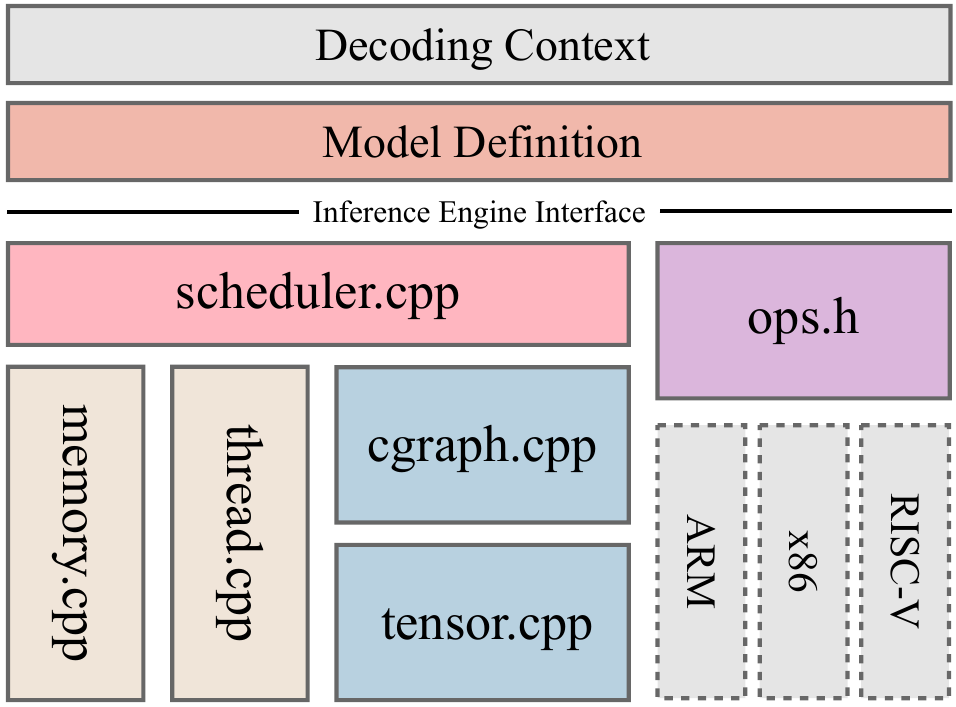}
  \caption{Architecture of \textsc{ArcLight}.}
  \label{fig:arch}
\end{figure}

\subsection{Overall Architecture}

As shown in Figure~\ref{fig:arch}, \textsc{ArcLight} adopts a decoupled architecture with a high-level decoding frontend and an underlying-level inference backend. This design keeps the core computation lightweight and largely hardware-agnostic, while enabling targeted optimizations in the engine.
The inference engine backend implements the minimal set of modules required for high-performance CPU inference. It comprises five core modules: memory manager, thread manager, tensor library, forward graph builder, and graph computation scheduler. Furthermore, the engine includes a specialized operator library that contains the mathematical kernels (e.g., GEMM, Softmax) required for LLM computations. The engine exposes a streamlined set of APIs to the frontend, which handles weight loading, model definition, and the autoregressive decoding loop. The entire architecture consists of about 10 C++ header files and source files.

\subsection{Tensor Library}

The tensor library defines the tensor and all operations of \textsc{ArcLight}, excluding its computation kernels. Unlike the C-style definition used in \textit{llama.cpp}, \textsc{ArcLight} adopts a strictly object-oriented encapsulation based on C++ classes to improve modularity and maintainability. An \textsc{ArcLight} tensor consists of two distinct components, i.e., header and data. The tensor header stores essential metadata, including the tensor name, shape, data type, operation type, auxiliary parameters, and pointers to source tensors for computation graph construction. The data area is a contiguous block of virtual memory. The library provides high-level interfaces to manage these attributes, such as methods to \textbf{get/set} names and shapes, or to calculate the total byte size required for memory allocation.

\subsection{Memory Manager}

\begin{figure}[t]
  \centering
  \includegraphics[width=1.00\columnwidth]{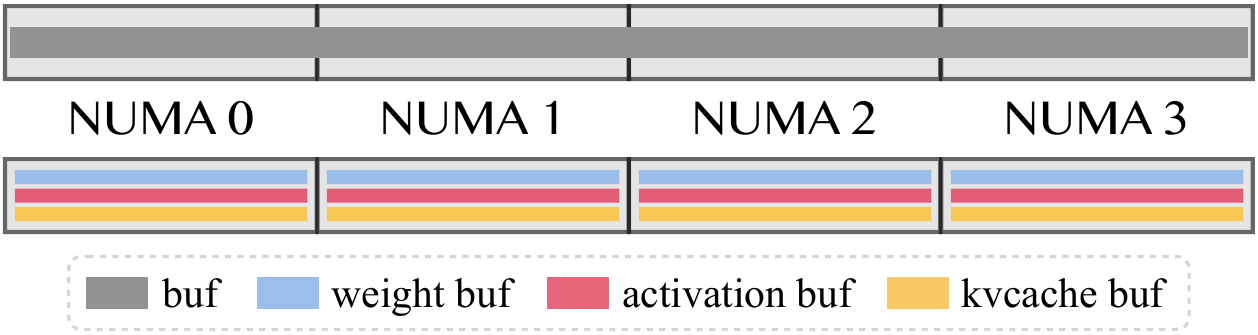}
  \caption{Comparison of UMA (top) and NUMA (bottom) buffer management strategy. In the UMA strategy, the NUMA nodes are transparent.}
  \label{fig:buffer}
\end{figure}

\begin{figure}[t]
  \centering
  \includegraphics[width=1.00\columnwidth]{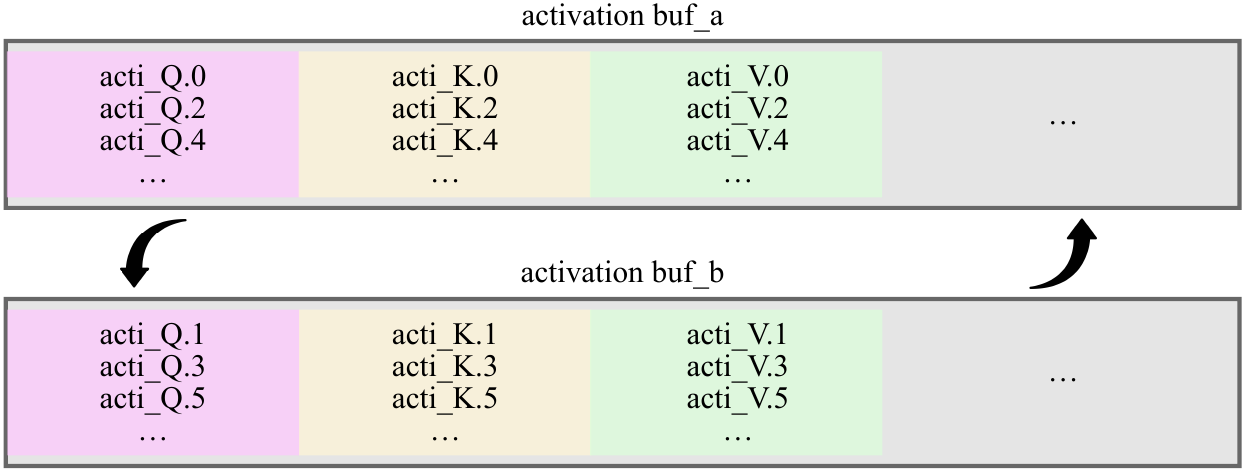}
  \caption{Double-buffering mechanism.}
  \label{fig:dual}
\end{figure}

The Memory Manager handles lifecycle management during runtime. Its primary responsibility is to pre-allocate a sufficient memory pool from the system at program startup, which is then allocated to weight and activation tensors. Different from the monolithic buffer (UMA) used in \textit{llama.cpp}, \textsc{ArcLight} allocates separate buffers in the local memory of each NUMA node when it is enabled, as illustrated in Figure~\ref{fig:buffer}. This design greatly simplifies explicit tensor-to-NUMA-node binding. Furthermore, to reduce memory overhead during layer-wise inference, we implement a double-buffering mechanism for the activation buffer. The two activation buffers are alternated based on layer parity, significantly lowering runtime memory consumption, as shown in Figure~\ref{fig:dual}.

\subsection{Thread Manager}

\begin{figure}[t]
  \centering
  \includegraphics[width=0.80\columnwidth]{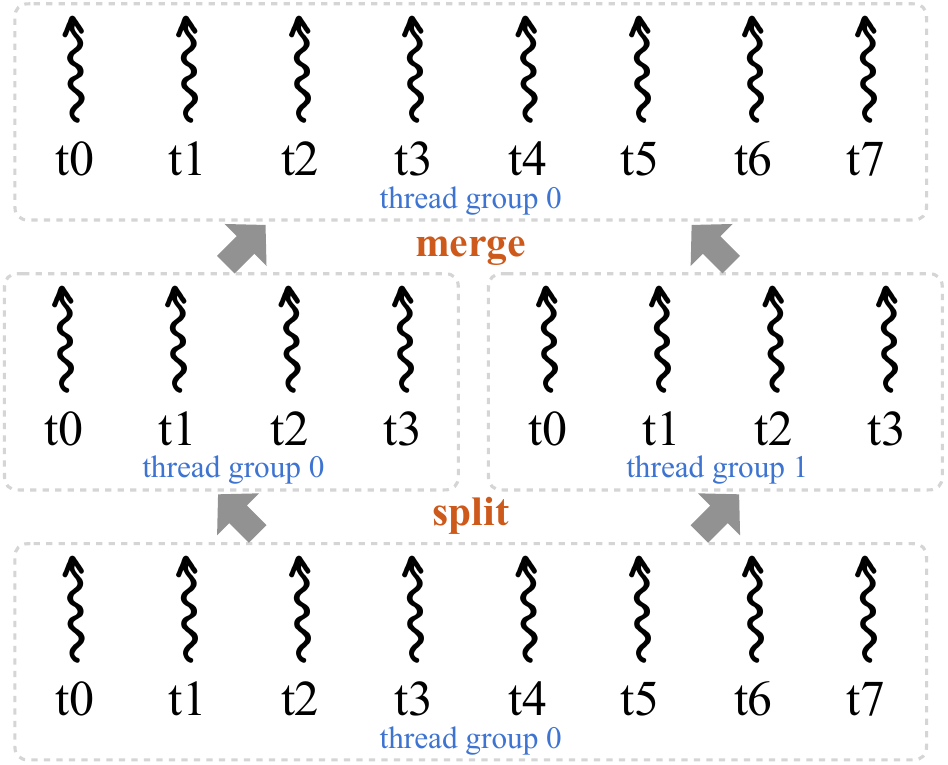}
  \caption{Multi-view thread organization.}
  \label{fig:thd}
\end{figure}

The thread manager creates a set of worker threads before inference begins, which collaboratively execute tensor computations during runtime. Common thread management like \textit{llama.cpp} organizes these workers into a single thread pool that executes the same computation task. However, this design is only suitable for a single forward computation graph. To flexibly support both single-graph and multi-subgraph execution, \textsc{ArcLight} introduces a multi-view thread organization, as illustrated in Figure~\ref{fig:thd}. We introduce the logical abstraction of thread groups within the thread pool. At both pool initialization and graph execution time, explicit interface and operators are provided to dynamically reconfigure the internal thread organization. When the pool is split into $n$ groups, these groups can execute $n$ independent tensor operations in parallel.

\begin{figure}[t]
  \centering
  \includegraphics[width=1.00\columnwidth]{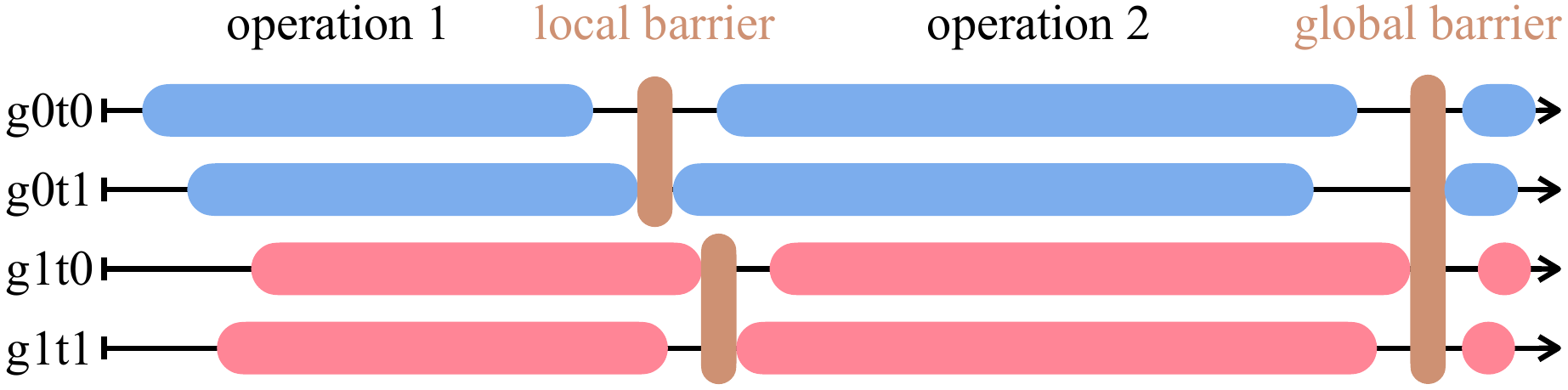}
  \caption{Local barrier and global barrier. ``g0t0'' denotes the 0-th thread in group 0.}
  \label{fig:barrier}
\end{figure}

\textbf{Barrier} synchronization among a group of threads is a common design in inference frameworks: all threads pass the barrier simultaneously to maintain consistent execution progress. When multiple logical thread organizations are introduced, barrier management becomes more complex. We therefore introduce a \textit{global barrier} for synchronization across the entire thread pool, as illustrated in Figure~\ref{fig:barrier}, to distinguish it from the legacy intra-group barrier. Without the global barrier, synchronization occurs only within each logical thread group. When enabled, threads across all logical groups must reach and pass the global barrier together.

\subsection{Forward Graph Builder}

Similar to \textit{llama.cpp}, \textsc{ArcLight} adopts a static computation graph, where the complete graph is constructed before execution. The graph builder implements and exposes a set of tensor operation interfaces that define graph nodes. These interfaces take source tensor pointers and necessary parameters as input, and return the output tensor corresponding to the current node. When defining a model in the frontend, one can construct the full computation graph simply by selecting and composing these interfaces.

In addition, we simplify the topological sorting procedure commonly used in mainstream frameworks. Since the model definition order naturally follows a topological order, we append each node to a sequential container at the end of its construction function, avoiding the cost of re-analyzing the graph structure. Detailed implementation is shown in Appendix~\ref{sec:graph}.

The computation graph module also implements KV cache management, including basic operations such as KV cache tensor creation, injection (set), and retrieval (get). Our graph construction design draws substantial implementation from \textit{llama.cpp}.

\subsection{Graph Computation Scheduler}

Once the computation graph is constructed, the execution order of all nodes is determined and stored in a sequential container. The scheduler iterates through this container in order, executing the computation associated with each node (e.g., GEMM, attention) before proceeding to the next. To ensure correctness and prevent interference across nodes, barrier synchronization is performed after the completion of each node’s computation. Computation graph construction and scheduling become more intricate under multi-NUMA optimization, which will be discussed in the next section.

\subsection{Tensor Operations}

The execution scheduler processes tensor operations in the graph sequentially. All operator interfaces are declared in \texttt{ops.h}, while their implementations depend on hardware features. Hardware-agnostic operations such as copy/reshape are organized in \texttt{common.cpp}. In contrast, CPU-dependent operations, including GEMM, FlashAttention, etc., are implemented according to the target hardware. For example, on ARM CPUs, different generations support varying levels of hardware acceleration for GEMM: NEON provides SIMD instructions, and i8mm accelerates 8-bit integer multiplication. \textsc{ArcLight} does not implement its own operator library. Instead, it reorganizes and leverages the operator implementations from \textit{llama.cpp}.

\section{Cross-NUMA Tensor Parallelism}

On many-core platforms, the latency gap between local memory access and cross-node access is substantial. Mainstream frameworks such as \textit{llama.cpp} do not explicitly optimize for this NUMA-induced memory barrier. Building upon the inference infrastructure described before, we introduce cross-NUMA tensor parallelism to overcome the performance limitations of many-core platforms.

\begin{table}[t]
\centering
\resizebox{0.8\columnwidth}{!}{
\renewcommand{\arraystretch}{0.9}
\begin{tabular}{c|cccc}
\toprule
\multirow{2}{*}[-0.6ex]{\begin{tabular}[c]{@{}c@{}}Cores\\from\end{tabular}} & \multicolumn{4}{c}{Memory in} \\ \cmidrule(lr){2-5}
 & node 0 & node 1 & node 2 & node 3 \\ 
\midrule
node 0 & 102 & 26 & 24 & 23 \\
node 1 & 26 & 103 & 23 & 22 \\
node 2 & 24 & 23 & 103 & 26 \\
node 3 & 23 & 22 & 26 & 101 \\
\bottomrule
\end{tabular}
}
\caption{The result of memory access speed (GB/s) of different core-memory combinations.}
\label{tab:numa}
\end{table}

\begin{figure}[t]
  \centering
  \includegraphics[width=0.95\columnwidth]{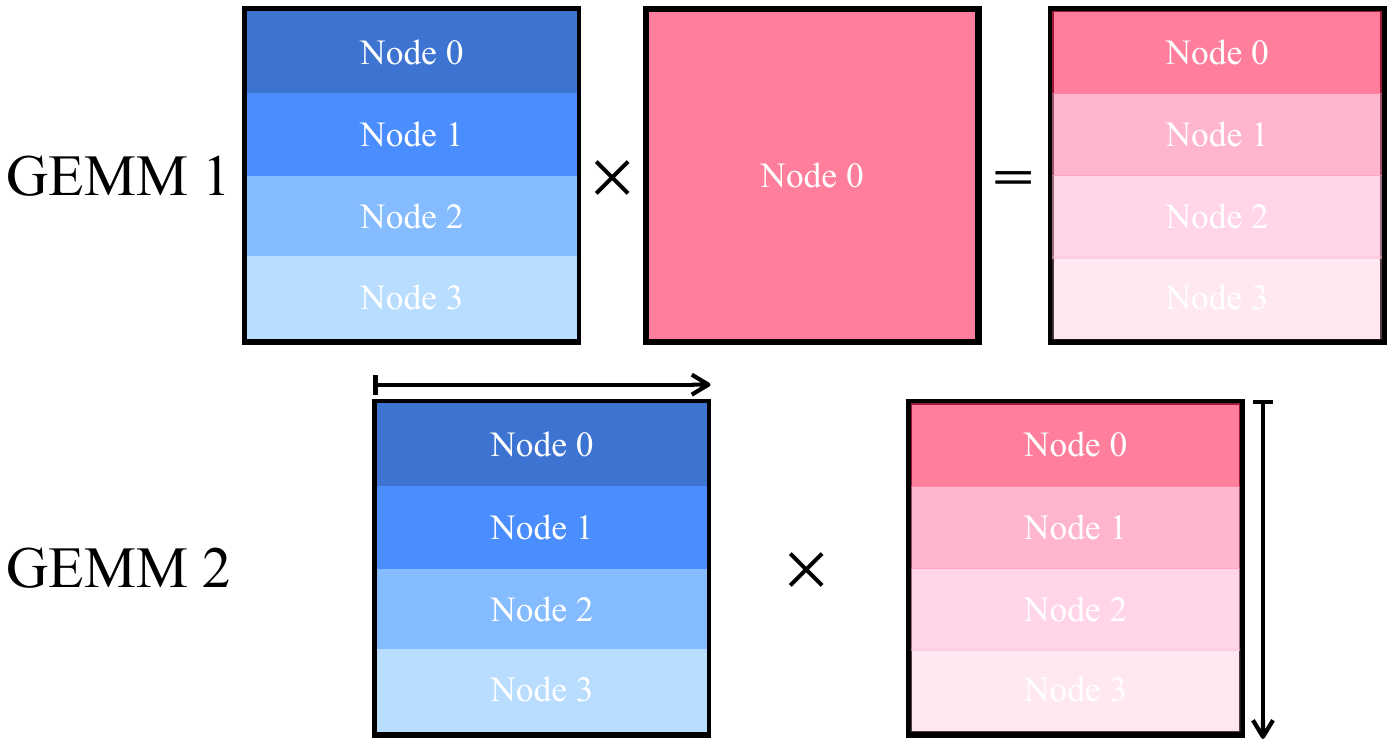}
  \caption{GEMM memory affinity in \textit{llama.cpp}. Blue denotes the weight, and red denotes the activation.}
  \label{fig:gemm}
  \vspace{-3mm}
\end{figure}

\subsection{Cross-NUMA Memory Access}

We first measure the memory access speed across different NUMA nodes. Using a 4-NUMA test machine, we benchmark memory access under different core–memory combinations and the results are reported in Table~\ref{tab:numa}. Each NUMA node on this system manages 6$\times$DDR4 memories and 48 ARM cores. The results show that cores accessing local memory are approximately 4 times faster than accessing remote memory across nodes. This clearly demonstrates the severe cross-node memory access wall on many-core platforms.

\begin{figure*}[t]
  \centering
  \includegraphics[width=0.85\textwidth]{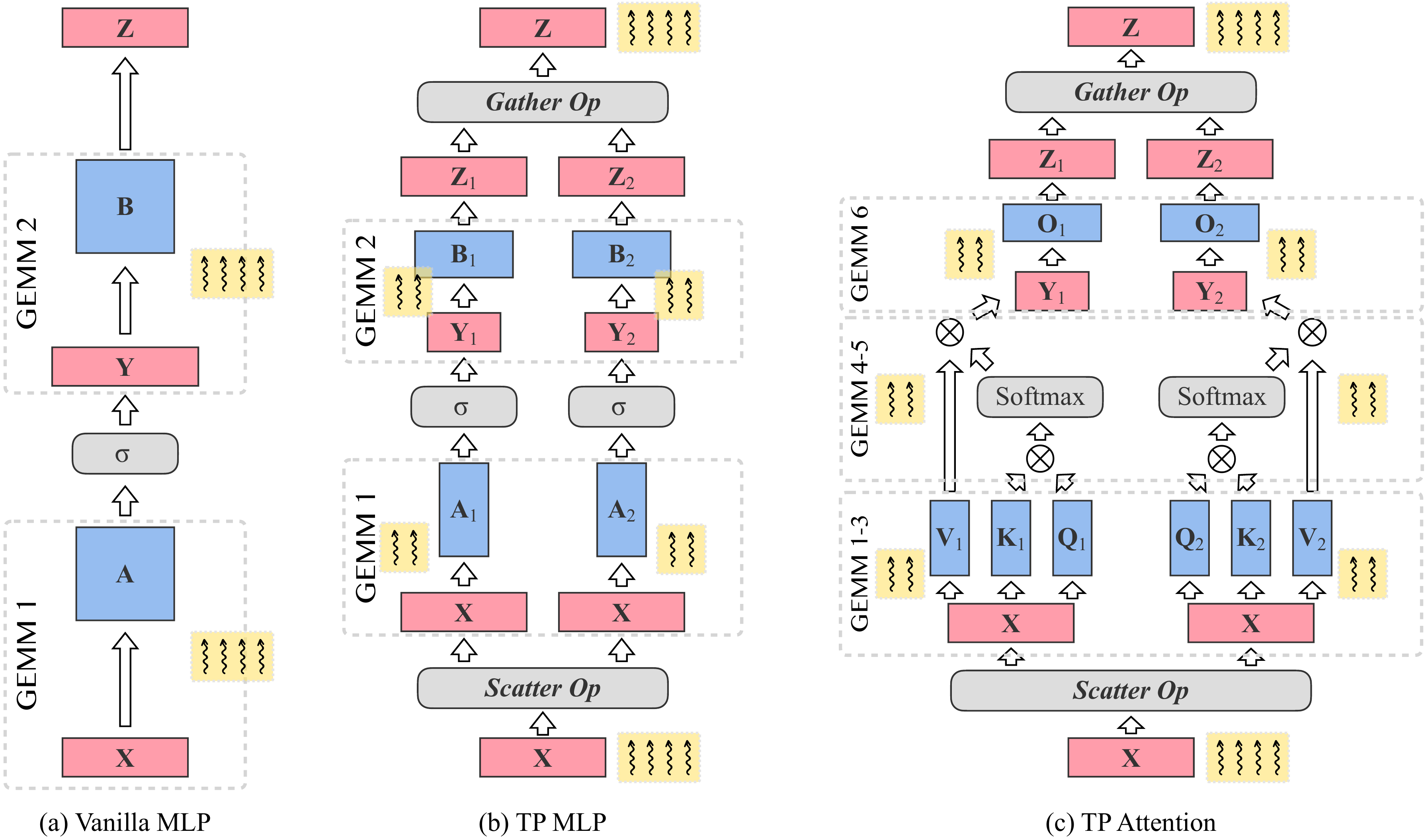}
  \caption{Simplified computation graph of vanilla MLP, tensor parallel MLP and Attention.}
  \label{fig:tp}
\end{figure*}

\begin{figure*}[t]
  \centering
  \includegraphics[width=0.90\textwidth]{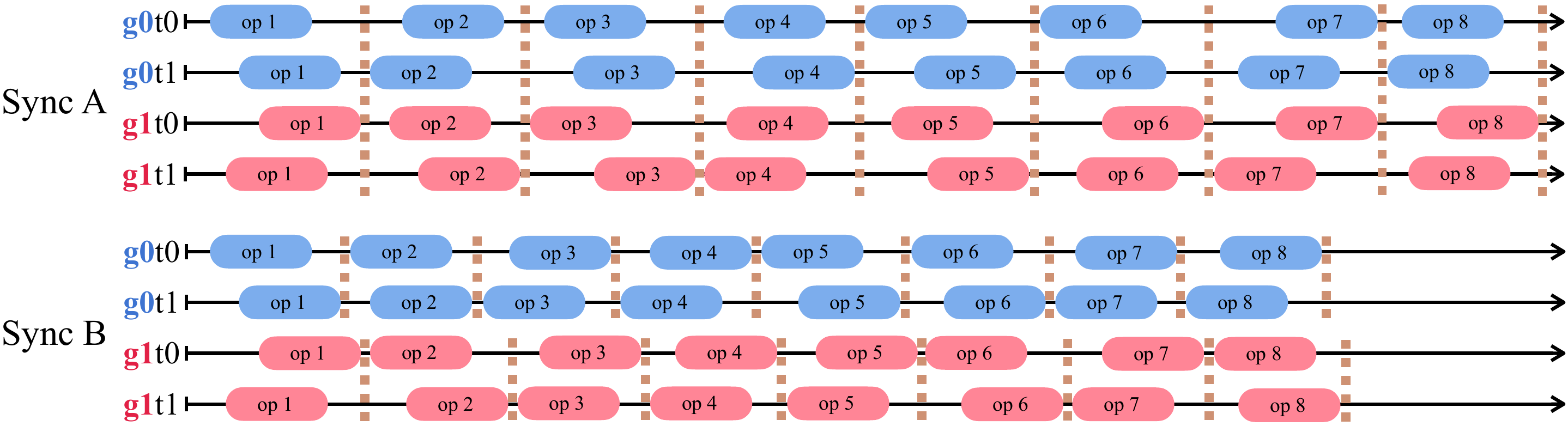}
  \caption{Comparison of different ways of thread group synchronization. The dashline represents a barrier. 
  }
  \label{fig:sync}
  \vspace{-2mm}
\end{figure*}

\textit{llama.cpp} provides only basic support for many-core platforms. It exposes a \texttt{--numa} option, and when set to \texttt{distribute}, the threads in its pool are evenly bound to cores across different NUMA nodes. However, \textit{llama.cpp} does not bind tensors to specific NUMA nodes, leading to frequent mismatches between computation and memory access. As shown in Figure~\ref{fig:gemm}, since threads are distributed across NUMA nodes, the corresponding weight areas are also partitioned accordingly. Following the first-touch policy, the operating system (OS) allocates physical memory pages for each weight partition on the NUMA node where it is first accessed, which is reasonable.

Consider GEMM1. Suppose the initial activation resides only on node 0. In this case, weight partitions on nodes 1–3 must access activations remotely, resulting in non-local memory access. After GEMM1, the output activation tensor becomes evenly distributed across all nodes. Now consider GEMM2. Although the weights remain partitioned across nodes 0–3, each weight partition finds that only one quarter of its activation accesses are local, while the remaining three quarters are remote. This pattern persists for subsequent GEMM operations, forming a sustained performance bottleneck.

\subsection{Weight Partition}

Tensor parallelism (TP) can partially reduce cross-node data access within consecutive GEMMs and is commonly used in multi-device LLM training~\citep{megatron2019}. When cross-NUMA memory access becomes a bottleneck on CPUs, TP can theoretically mitigate its impact. As shown in Figures~\ref{fig:tp}a and~\ref{fig:tp}b, compared with a vanilla MLP, a properly partitioned sequence of matrix multiplications eliminates cross-node memory access. In transformer models, $\mathbf{W}_q, \mathbf{W}_k, \mathbf{W}_v, \mathbf{W}_{gate}$, and $\mathbf{W}_{up}$ are row-partitioned, while $\mathbf{W}_o$ and $\mathbf{W}_{down}$ are column-partitioned. In other words, $\mathbf{W}_q, \mathbf{W}_k$, and $\mathbf{W}_v$ are partitioned by attention heads. Taking the MLP in the figure as an example, the original computation is $\mathbf{Y} = \sigma(\mathbf{AX})$, $\mathbf{Z} = \mathbf{BY}$.
Under TP, it becomes $[\mathbf{Y}_1, \mathbf{Y}_2] = [\sigma(\mathbf{A}_1 \mathbf{X}), \sigma(\mathbf{A}_2 \mathbf{X})]$, $
[\mathbf{Z}_1, \mathbf{Z}_2] = [\mathbf{B}_1 \mathbf{Y}_1, \mathbf{B}_2 \mathbf{Y}_2]$,
and $\mathbf{Z} = \mathbf{Z}_1 + \mathbf{Z}_2$. All tensors involved in TP are split into buffers under each NUMA node, thus effectively isolating cross-node memory access.

\subsection{Parallel Computation Graph}

Without TP, all threads cooperatively execute \textbf{the same} operation (e.g., GEMM). After introducing TP on CPUs, multiple operators can run concurrently. Consequently, the computation graph becomes a set of parallel subgraphs, requiring flexible switching between single-graph and multi-subgraph execution, as shown in Figure~\ref{fig:tp}. We implement two core operators: \textit{Scatter} and \textit{Gather}. The Scatter operator reconfigures the thread pool into multiple groups and creates view tensors for the input of each subgraph. Once Scatter completes, the subgraphs execute in parallel. The Gather operator collects and sums the output tensors from all subgraphs, and restores the thread pool to a single group. After Gather finishes, execution returns to non-TP mode.

\subsection{Thread Synchronization}
\label{sec:async}

As discussed earlier, once execution enters multi-subgraph TP mode, the subgraphs are mutually independent and have no data dependencies. Intuitively, they execute asynchronously, even when performing the same GEMM concurrently. When TP is enabled, synchronization across thread groups can follow two ways, as shown in Figure~\ref{fig:sync}. In Sync A, all thread groups use global synchronization: every group completes one operator before jointly proceeding to the next. In Sync B, synchronization is confined within each thread group, while global barriers are applied only at the beginning and the end. Both the figure and empirical results show that \textbf{asynchronous subgraph execution} significantly reduces thread idle time, further accelerating TP execution.

\section{Experiments}


We evaluate the decoding throughput of \textsc{ArcLight} against llama.cpp on a 192-core test machine with 4 NUMA nodes. Each node contains 48$\times$HUAWEI Kunpeng-920 cores (ARMv8.2) and 6$\times$DDR4 memory channels. Note that \textsc{ArcLight} is compatible with widely available ARM CPUs. The OS is Ubuntu 22.04. The prompt length is 15 tokens, and the generation length is 256 tokens. The results for a prompt length of 300 tokens and the corresponding prefill throughput performance are provided in Appendix~\ref{sec:speed}. Tensor operations use the \texttt{NEON} vector instructions. We use Qwen3-4B~\citep{yang2025qwen3} for our experiments, and it is quantized in the Q4\_0 format.

\begin{figure}[t]
  \centering
  \includegraphics[width=0.80\columnwidth]{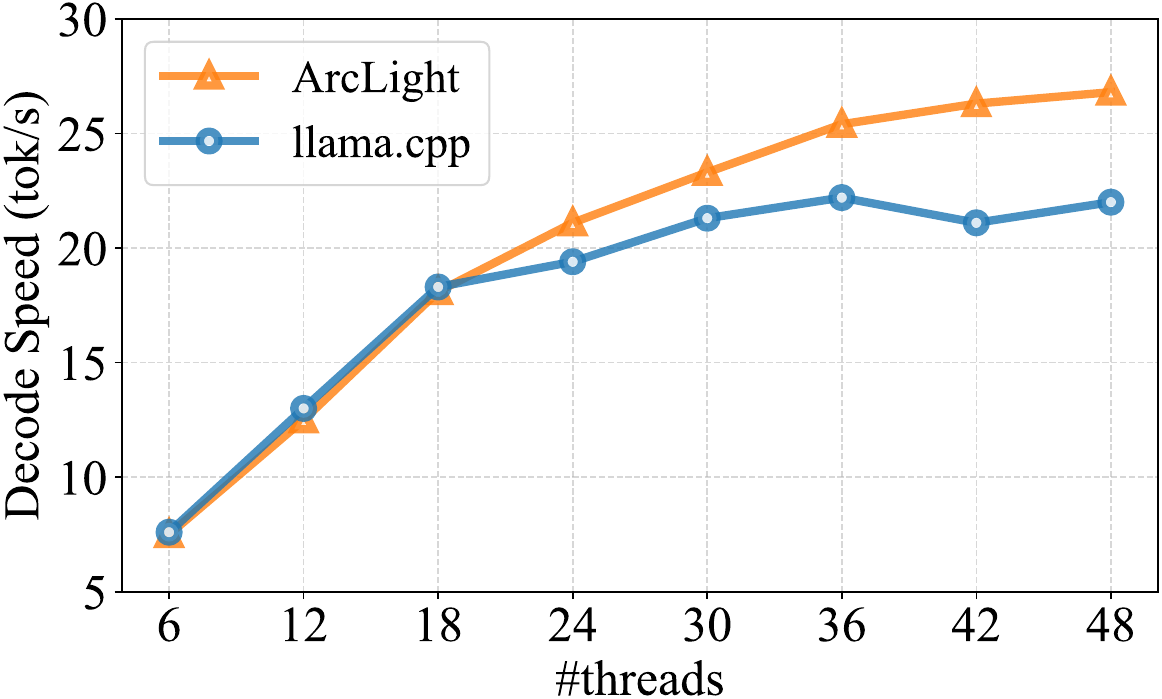}
  \caption{Decoding speed using a single NUMA node.}
  \label{fig:base}
\end{figure}

We first compare the case where all threads are bound to a \textbf{single NUMA node}, as shown in Figure~\ref{fig:base}. As the number of inference threads increases from 6 to 48, the decoding throughput of both frameworks improves. This indicates that, with high intra-node memory bandwidth, throughput performance scales with the number of cores. Because the buffer in \textit{llama.cpp} is allocated without NUMA-aware placement and its pages are distributed by the OS across nodes, whereas \textsc{ArcLight} enforces node-local allocation, \textsc{ArcLight} achieves slightly higher throughput.

\begin{figure}[t]
  \centering
  \includegraphics[width=0.80\columnwidth]{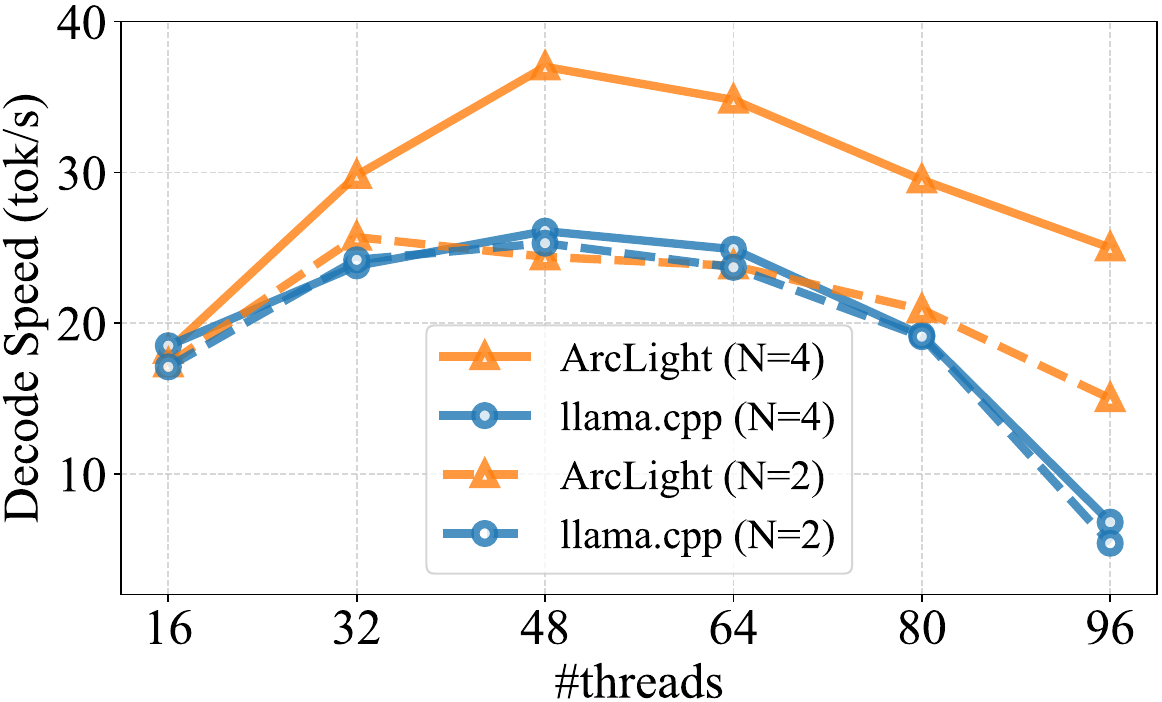}
  \caption{Decoding speed using multiple NUMA nodes. ``N'' is the number of NUMA nodes.}
  \label{fig:decode_10}
\end{figure}

We next evaluate the case where threads are evenly distributed across \textbf{multiple NUMA nodes}, as shown in Figure~\ref{fig:decode_10}. The results are reported for both 2-node and 4-node cases. \textsc{ArcLight} with TP outperforms \textit{llama.cpp}, which only distributes thread bindings while leaving memory placement to the OS. The improvement comes from breaking the single-node memory access wall through cross-NUMA TP. Notably, the asynchronous subgraph execution introduced in Section~\ref{sec:async} contributes an additional gain of about 5 token/s.

\section{Related Work}

There are many popular LLM inference frameworks, such as vLLM~\citep{efficient2023} and TensorRT~\citep{tensorrt2024} on GPUs, and llama.cpp~\citep{llamacpp2023} and ONNX Runtime~\citep{onnx2024} on CPUs. These frameworks address device-specific bottlenecks in memory management and data access using techniques such as FlashAttention~\citep{flashattention2023}, Paged Attention~\citep{efficient2023} , Pipeline Parallelism~\citep{gpipe2019}, and Sequence Parallelism~\citep{reducing2023}. However, they largely overlook the cross-NUMA bottleneck on many-core CPU platforms.

The most common way to address memory bandwidth limitations during inference is model compression, such as through 2-bit~\citep{omniquant2023,quip2024} or 1-bit~\citep{onebit2024xu,xu2025crvq} quantization and pruning~\citep{wanda2024,xu2025camera}. In addition, overcoming the bandwidth upperbound of a single node through appropriate parallelism is also a viable approach~\cite{megatron2019,reducing2023}. Most of these techniques become key acceleration methods in mainstream inference frameworks.

\section{Conclusion}

We implement a lightweight LLM inference architecture on CPU with redesigned memory and thread management, and computation graph execution, while introducing targeted optimizations for many-core platforms. Compared with its mainstream counterpart, \textsc{ArcLight} achieves substantial performance improvements. We will long-term maintain and open development of it, with the goal of establishing it as an accessible, practical development toolkit and educational platform.

\section*{Limitations}


Despite our efforts toward efficient CPU inference, several limitations remain. \textsc{ArcLight} is currently evaluated only on ARM platforms, and support for additional architectures (e.g., x86) is left for future work. In addition, our implementations of Scatter and Gather operators are still preliminary. Further optimization will focus on reduce memory overhead and improve parallel efficiency.

\section*{Acknowledgments}

We gratefully acknowledge the support of the National Natural Science Foundation of China (NSFC) via grant 62236004, 62476073 and 62576186. This work is also supported by the National Key Research and Development Program of China (2024YFB4505603) and the high-quality development project of MIIT.


\bibliography{custom}

\appendix

\section{Appendix}
\label{sec:appendix}

\subsection{Forward Graph Builder}
\label{sec:graph}

The forward graph builder provides modular interfaces required by LLMs, chaining graph nodes in a compositional manner. In legacy inference frameworks, these interfaces take source tensor pointers and necessary parameters as input, and return a tensor pointer for the current module. For example, a Linear layer accepts weight and input activation tensor pointers and returns the output tensor pointer. With tensor parallelism (TP), the computation graph contains subgraphs that execute in parallel. The framework must satisfy two requirements: (1) preserve a unified module interface so that model definitions do not need to be rewritten for a TP-specific version; and (2) support construction of parallel computation graphs. To achieve this, we extend the tensor pointer type \texttt{tensor*} to a tensor bundle type \texttt{tensor\_ptrs}, which maintains a set of tensor pointers and supports mutual assignment with a single tensor pointer. By replacing the original tensor pointer type in module interfaces with \texttt{tensor\_ptrs}, we enable seamless reuse of existing interfaces when TP is enabled.

As described in the main text, we simplify the topological sorting process from computation graph construction to execution order generation. \textsc{ArcLight} uses a static linked list (array-based) as a sequential container to maintain the execution order of graph nodes. At the end of each module construction function, the current tensor bundle is appended to the tail of this list. Each node also stores the index of its successor. The static list supports four construction modes.
\begin{itemize}
    \item \textbf{Serial mode}: the conventional append operation, where each tensor bundle contains a single tensor and is added to the tail.
    \item \textbf{Scatter mode}: a tensor bundle containing multiple tensors is appended to a single tensor pointer, enabling the transition from a single graph to multiple parallel subgraphs.
    \item \textbf{Parallel mode}: within TP-enabled modules, each tensor in a bundle is appended one-to-one to the corresponding tensor in the previous bundle.
    \item \textbf{Gather mode}: a single tensor is appended to a multi-tensor bundle at the tail, enabling the transition from multiple subgraphs back to a single graph.
\end{itemize}

\begin{figure}[t]
  \centering
  \includegraphics[width=0.80\columnwidth]{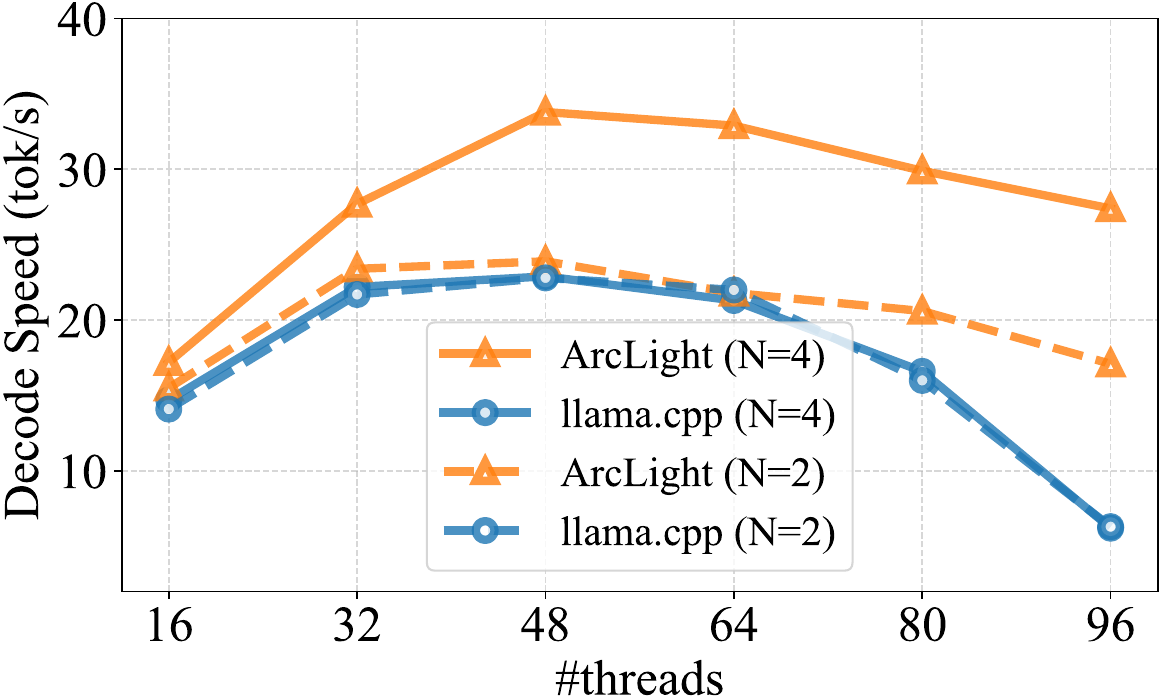}
  \caption{Decoding speed using multiple NUMA node when the length of prompt equals 300. ``N'' is the number of NUMA nodes.}
  \label{fig:decode_300}
\end{figure}

\begin{figure}[t]
  \centering
  \includegraphics[width=0.80\columnwidth]{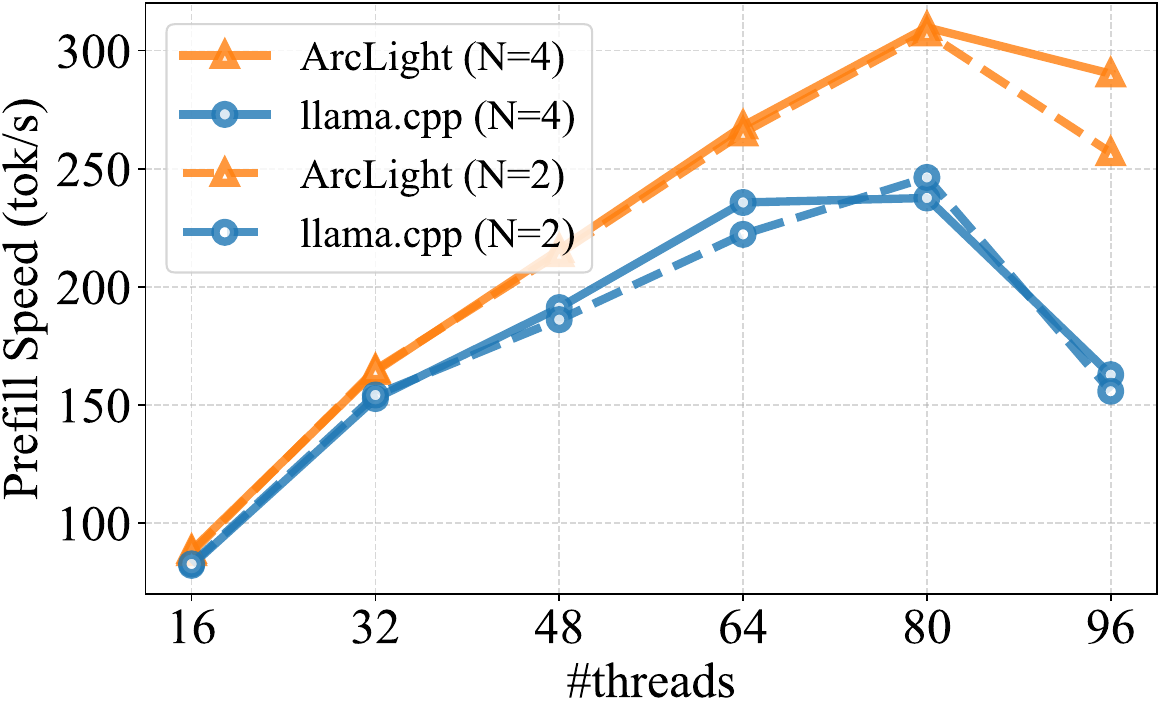}
  \caption{Prefilling speed using multiple NUMA node when the length of prompt equals 300. ``N'' is the number of NUMA nodes.}
  \label{fig:prefill_300}
\end{figure}

\subsection{Supplementary Speed Test}
\label{sec:speed}

We report prefilling and decoding throughput with a prompt length of 300. All experiments are conducted under multi-NUMA configurations, with other settings identical to those in the main text. As shown in Figure~\ref{fig:decode_300}, decoding throughput is slightly lower than that with short prompts. Moreover, although \textsc{ArcLight} still outperforms \textit{llama.cpp} in prefill throughput, the advantage is less pronounced than during decoding, as shown in Figure~\ref{fig:prefill_300}. This is because TP in \textsc{ArcLight} primarily addresses the memory access wall, whereas prefill is predominantly compute-bound.

\subsection{Test Commands for llama.cpp}
\label{sec:command}

We provide the command used to launch \textit{llama.cpp} for benchmarking reference. For the single-NUMA-node evaluation, the test can be launched with the following command:

{\small
\begin{tcolorbox}
\texttt{llama-cli -m Qwen3-4B-Q4\_0.gguf --threads 48 -ctk q4\_0 -ctv q4\_0 -n 256 --top-k 1 --numa isolate}
\end{tcolorbox}
}

For the multi-NUMA evaluation, if threads are to be evenly distributed across all NUMA nodes, the benchmark can be launched as follows:

{\small
\begin{tcolorbox}
\texttt{llama-cli -m Qwen3-4B-Q4\_0.gguf --threads 64 -ctk q4\_0 -ctv q4\_0 -n 256 --top-k 1 --numa distribute}
\end{tcolorbox}
}

\end{document}